\documentclass[aps,pra,twocolumn,showpacs,groupedaddress,superscriptaddress,amsmath,amssymb]{revtex4}
\usepackage{graphicx}
\usepackage{natbib}

\begin{document}


\title{Observation of polarization quantum noise of laser radiation in Rb vapor cell
}
\author{A. S. Zibrov}
\affiliation{Department of Physics, Harvard Univ., Cambridge, MA,
02138, USA } \affiliation{Harvard-Smithsonian Center for
Astrophysics, Cambridge, MA, 02138, USA } \affiliation{Lebedev
Institute of Physics, Moscow, 117924, Russia}

\author{I. Novikova}
\affiliation{Harvard-Smithsonian Center for Astrophysics, Cambridge,
MA, 02138, USA }

\begin{abstract}
We present experimental study of polarization quantum noise of laser
radiation passed through optically think vapor of Rb$^{87}$. We
observe a step-like noise spectrum. We discuss various factor which
may result in such noise spectrum and prevent observation of
squeezing of quantum fluctuations predicted in Matsko et al, Phys.\
Rev.\ A ${\bf63}$, 043814 (2001).
\end{abstract}

\date{\today}
\maketitle

The sensitivity of many experiments is ultimately limited by quantum
fluctuations of electromagnetic field. This stimulates the
development of various methods for generation of light with
non-classical statistics (``squeezed light''). Usage of such light
field in an experiment often allows measurement noise reduction below
shot noise limit~\cite{Scully-Book,Bachor}.
%
%
Photon statistics modification often requires nonlinear optical
medium; the degree of change depends on the strength of the nonlinear
effects and linear losses in such medium~\cite{Bachor,JOSASqueeze}.

Recent proposals of photon statistics control based on coherent
atomic vapor look quite promising from that point of view. Atomic
coherence created between two hyperfine ground states of an alkali
metal by a strong classical control field eliminates linear
absorption and simultaneously produces strong nonlinear dispersion
for a weak probe field (classical or quantized). This effect is known
as electromagnetically induced transparency (EIT)~\cite{Scully-Book}.
Third-order nonlinear susceptibility in an EIT medium is substantial
and comparable with linear susceptibility, which enhances coupling
between electromagnetic fields participating in such nonlinear
processes~\cite{harris}.
%
Ref.~\cite{Caspar} demonstrated effective projection of photon
quantum state on collective spin excitation of an atomic ensemble
under EIT condition, long storage time, and the on-demand retrieval
of quantum information.
%

Matsko \textit{et al.}~\cite{Matsko} predicted squeezing of vacuum
quantum fluctuations for laser light propagating through an EIT
medium with coherence between Zeeman states of the same hyperfine
atomic sublevel. Theoretical calculations estimated squeezing up to
8dB under realistic experimental conditions.
%

The main purpose of this work was to study the polarization quantum
noise of laser radiation for the $D_1$ line of Rb and to explore the
possibility for squeezed vacuum observation on that transition. In
particular, we studied the relation between laser intensity and the
bandwidth of laser-induced quantum noise.

\begin{figure}[ht]
\includegraphics[width=6cm]{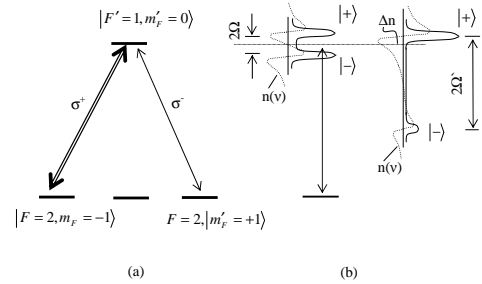}
\caption {\label{fig1} a) $\Lambda$ system formed on Zeeman sublevels
by the circularly polarized components $\sigma^+$ and $\sigma^-$ of
the linearly polarized laser field. b)~Same $\Lambda$ system in the
dressed-state basis for stationary atoms (left part) and the atoms
moving with the speed $\kappa\upsilon$ (right part). $\Omega$ is the
Rabi frequency of the laser field,
$\Omega^\prime=\sqrt{\Omega^2+(\kappa\upsilon)^2}$ , $n(\nu)$ is the
refractive index dispersion. }
\end{figure}

The mechanism of polarization quantum noise squeezing proposed
in~\cite{Matsko} is the following: linearly polarized laser radiation
is decomposed into two circularly polarized components $\sigma^+$ and
$\sigma^-$, as shown in Fig.~\ref{fig1}(a), which form a $\Lambda$
system and create coherence between $|m_F=-1\rangle$ and
$|m_F=+1\rangle$ Zeeman sublevels, producing EIT. In an ideal
three-level $\Lambda$ system phase difference for two circularly
polarized components does not change after interaction with atoms.
However, under realistic conditions interaction of light with
additional atomic levels (in particular with the other far-detuned
hyperfine sublevel(s) of the excited state), Doppler effect, optical
losses, etc., causes some changes in the relative phase between the
$\sigma^+$ and $\sigma^-$ components.
Indeed, in a dressed-state basis~\cite{Cohen} the absorption spectrum
of $\sigma^-$ consists of two transitions to $|+\rangle$ and
$|-\rangle$ states with corresponding spectral distribution of
refractive index $n(\nu)$. Emission of a spontaneous photon at a
frequency $\nu_0+(\kappa\upsilon)$ changes the shift $2\Omega^\prime$
between dress states $|+\rangle$ and $|-\rangle$ (as shown in
Fig.~\ref{fig1}(b)), which affects the transmission of the $\sigma^-$
field. Simultaneously, resonance refractive index changes by $\Delta
n$, creating correlation between fluctuations of the photon number
and the phase of the light field due to ac-Stark effect.
%
%
As a result quantum noise of one laser field quadrature can be
reduced at the expense of the other.


Experiments by Ries \textit{et al.}~\cite{Lvov} confirmed the
theoretical proposal of Matsko \textit{et al.}~\cite{Matsko} ,
reporting detection of squeezed vacuum at the output of a Rb vapor
cell for linearly polarized light resonant with the $D_2$ transition.
Observed reduction of a quadrature noise below standard quantum noise
limit was $-0.85$~dB. It is reasonable to assume that the $D_1$ line
is more promising for effective vacuum squeezing generation, since
the $D_1$ line has simpler excited level structure with all sublevels
participating in the coherence formation, and therefore exhibits
stronger EIT and nonlinear magneto-optical effects~\cite{Novikova}.

%
%

\begin{figure}[ht]
\includegraphics[width=8cm]{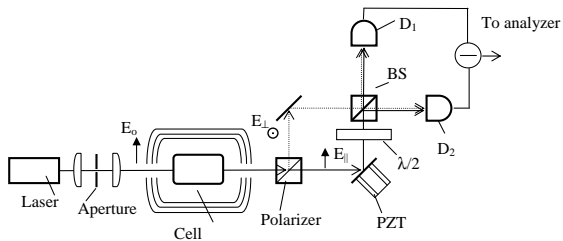}
\caption {\label{fig1a} Schematic of the experimental setup. $E_{\|}$
marks the local oscillator channel. Orthogonally polarized radiation
$E_{\bot}$ propagated in the other interferometer channel. Relative
phase of two interferometer channels was controlled by the mirror
mounted of the piezo-drive. Polarization of the local oscillator
field was rotated by $90^\circ$ using a half-wave plate to observe an
interference with the vacuum field.
}
\end{figure}

A schematic of the experimental setup is shown in Fig.~\ref{fig1a}.
Linearly polarized laser radiation propagated through a cylindrical
Pyrex vapor cell containing isotopically enriched $^{87}$Rb (cell's
length and diameter were $2.5$~cm). The cell was mounted inside a
three-layer magnetic shield and heated up to $\sim 95^\circ$С, which
corresponds to a Rb density of N = $2\times
10^{12}~\mathrm{cm}^{-3}$. Frequency of the external cavity diode
laser was close to $F=2 \rightarrow F'=1,2$ of the $D_1$ line of
$^{87}Rb$ (wavelength $\lambda=795$~nm). Laser power before the cell
was $P=1 \div 8$~mW; laser beam diameter was $D = 1$~mm. To reduce
asymmetry of the diode laser transverse intensity distribution we
used a spatial filtering focusing the laser bean onto a $30~\mu$m
pinhole. This system provided nearly Gaussian transverse intensity
distribution with $\sim$70\% transmission.
%
%
We used a traditional phase-sensitive homodyne detection
scheme~\cite{Mandel}, which included a Mach-Zehnder interferometer
formed by a crystal polarizer with extinction ratio of $5\times
10^{-6}$ to separate laser field $E_{\|}$ and orthogonally-polarized
squeezed vacuum field $E_{\bot}$, mirrors, a half-wave plate to
adjust the polarization of the local oscillator, and a 50:50
non-polarizing beam splitter (BS). The strong linearly polarized
laser field $E_{\|}$ played the role of a local oscillator.
%

To align the interferometer we inserted a quarter-wave plate
$\lambda$/4 before the polarizer, sending the same intensity to the
both interferometer channels. Best fringes visibility was
$\sim$(96-98)\%, which is evidence of a single-mode laser field.
%

Light at the outputs of the homodyne detection scheme was collected
at two identical silicon $p-\iota-n $ photodetectors
$\mathrm{D}_{1,2}$ with quantum efficiency of 91\%(Hamamatsu S3883).
The two photocurrents were amplified using a low-noise amplifiers
(OPA657) and subtracted using a $180^\circ$ combiner (MiniCircuits
ZFSCJ-2-2) with $0.01 - 20.0$~MHz. We carefully balanced the
amplification in the two inputs. We modulated the laser current at
$\sim5$~MHz frequency, which produced a corresponding peak in the
laser spectrum. Then for equal laser intensities in the two
interferometer channels we adjusted the amplification of the
photodiodes such that the $5$~MHz peak disappeared after the
photocurrent subtraction. This procedure provided the accuracy in
photocurrent subtraction better than $35$~dB.
%

To measure a standard quantum noise level (denoted as SQL) for the
homodyne detector we blocked a vacuum channel $E_{\bot}$ and measured
a noise spectrum of the subtracted photocurrents as a function of the
laser intensity. We observed a linear dependence for a whole range of
the laser intensities used in the experiment. Blocking one of the
photodiode reduced the detected noise level by $\sim 1.4$.
%

\begin{figure}[ht]
\includegraphics[width=8cm]{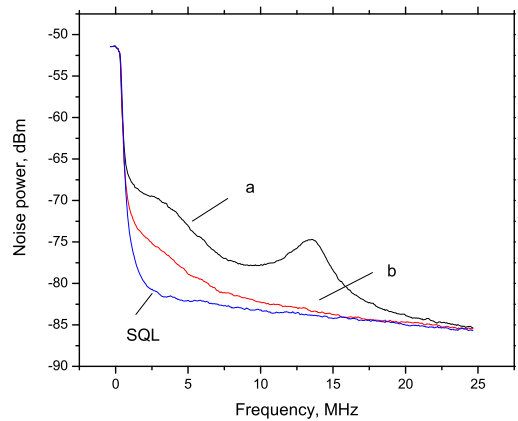}%
\vfill
 \caption {\label{fig2}
Homodyne detector noise spectrum when the relative phase between two
interferometer channels is $0^\circ$ (a) and $180^\circ$ (b). SQL -
standard quantum limit - is the noise spectrum with the vacuum
interferometer channel $E_{\bot}$ blocked. Laser power is $7.4$~mW.
Laser frequency is blue-detuned by $150$~MHz from $F=2\rightarrow
F'=1$ transition. Spectrum analyzer registration bandwidth is
$100$~kHz, video bandwidth is $30$~Hz. }
%
\end{figure}

\begin{figure}[ht]
\includegraphics[width=8cm]{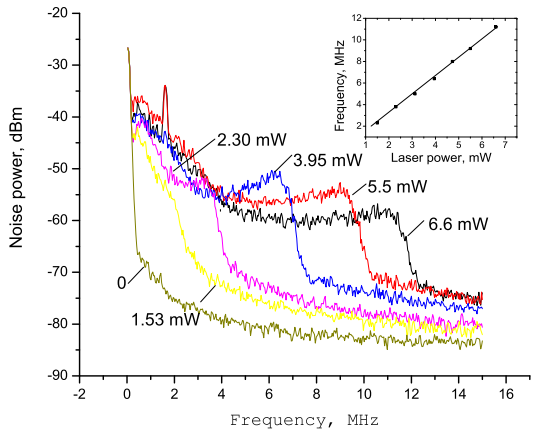}%
\vfill
 \caption {\label{fig3}
Homodyne detector noise spectra for different laser power for the
constant zero relative phase between two interferometer channels.
Laser frequency is blue-detuned by $150$~MHz from $F=2\rightarrow
F'=1$ transition. Spectrum analyzer registration bandwidth is
$100$~kHz, video bandwidth is $1$~kHz.
\newline\textit{Inset}: Dependence of the noise ``step'' position on laser power.}
\end{figure}

We studied the quantum noise properties of the laser radiation tuned
near the Doppler-broadened (HWHM $\approx 400$~MHz) $F=2\rightarrow
F'=1,2$ transitions of ${87}$Rb $D_1$ line (excited state hyperfine
splitting is $812$~MHz). We observed enhancement of the noise both
below and above the resonance frequencies. For the detuning
$\sim(+150 \pm 60)$~MHz from $F=2\rightarrow F'=1$ transition the
shape of the noise spectrum looked like a step with a peak. A sample
of such a spectrum is shown in Fig.~\ref{fig2}. The position of the
step was proportional to the laser intensity, and it shifted to the
higher frequencies at more intense laser field (see Fig.~\ref{fig3}).
Such spectral dependence was observed only for one noise quadrature.
If the relative phase between two interferometer channels changed by
$180\circ$, which corresponds to the orthogonal quadrature detection,
the noise level dropped by $10-20$~dB, as shown in Fig.~\ref{fig2}b.
%
%
We failed to observe any vacuum squeezing in the experiment, as the
the noise of the homodyne detector never dropped below the standard
quantum level. The minimum excess noise occurred for the spectral
frequency range close to the laser field Rabi frequency $\Omega$
(e.g., the Rabi frequency for the Fig.~\ref{fig2}a was $\Omega \sim
26$~MHz). In the previous experiments~\cite{Lvov} the bandwidth of
the detected vacuum squeezing was close to $5$~MHz while the Rabi
frequency was $\sim100$~MHz.

%

\textbf{Discussion} \\Let us first point out the difference between
experimental conditions in this work and the previous publications.
In Refs.~\cite{Matsko, Lvov} vacuum squeezing was predicted/observed
for very high laser intensity comparable with hyperfine splitting of
the ${}^{87}$Rb ground state ($\sim 6.8$~GHz). Under such conditions
the influence of atomic coherence on nonlinear properties of an
atomic vapor is insignificant~\cite{Matsko2}. Under such conditions
the $D_1$ transition of ${}^{87}$Rb can be treated as a two-level
system $J=1/2\rightarrow J'=1/2$, where nonlinear circular
birefringence occurs due to optical pumping. Resonance absorption in
such a system changes faster than nonlinear dispersion, which allowed
Matsko \textit{et al.} to find an optimal laser detuning to observe
quadrature squeezing of vacuum fluctuations.
%
%
In the present experiment atomic coherence was the leading mechanism
for nonlinear circular
birefringence~\cite{novikova02jmo,novikova'00ol}, and the EIT
parameters defined the quantum noise spectrum . For example,
correlations between intensity and phase fluctuations were the most
pronounced near the sharp boundary of the transparency window which
occurred in optically dense EIT media. All noise components outside
of the transparency window $\Gamma$ are
absorbed~\cite{Caspar,Zhenia,Zibrov-Dense}. EIT resonance width in an
optically dense atomic vapor is proportional to the light
intensity~\cite{Zibrov-Dense}:
%

\begin{equation}
\Gamma =
\frac{\Omega^2}{\sqrt{\gamma_{bc}{\gamma_a}}}\frac{1}{\sqrt{\eta\kappa
L }} \label{eq2}
\end{equation}


where $\Omega$ is the laser field Rabi frequency, $\gamma_a$ is the
dephasing rate of the optical transition, $\gamma_{bc}$ is the
ground-state decoherence rate, $\eta=3N\lambda^3/4\pi^2$, $N$ is
atomic density, and $\kappa=2\pi/\lambda$.

\begin{figure}[ht]
\includegraphics[width=6cm]{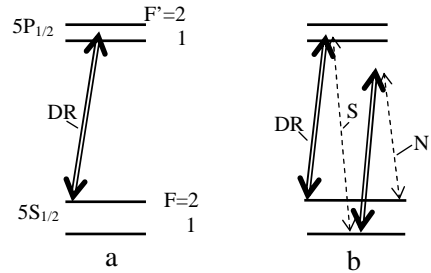}
 \caption {\label{fig4}
a) ${}^{87}$Rb $D_1$ line interacting with a strong drive field $DR$.
b) Spontaneous Raman scattering, producing two new fields $S, N$ (see
\cite{Zibrov-Parametric} for details).}
\end{figure}

Let us estimate the value of $\Gamma$. In a vapor cell $\gamma_{bc}$
is inversely proportional to the average time-of-flight of thermal Rb
atoms through the laser beam, and is approximately equal to
$\gamma_{bc}\approx 10^5$~Hz; the radiative decay of the excited
state is $\gamma_a = \sim 6$~MHz. We can estimate Rabi frequency
using the following expression~\cite{Maleki}:
$\Omega=\gamma_{а}\sqrt{I/8}$, where $I$ is the laser intensity in
$(\mathrm{mW/cm^{2}})$. Laser power $\approx 7$~mW gives a Rabi
frequency $\Omega\approx 25$~MHz, and $\Gamma \sim 20$~MHz. This is
relatively close to peak position in the noise spectrum shown in
Fig.~\ref{fig2}. The discrepancy is due to the inhomogeneous
transverse laser intensity distribution and other factors~\cite{Ye}.


An optically dense coherent atomic vapor is known to enhance
spontaneous Raman process shown in Fig.~\ref{fig4}, resulting in
generation of new Stokes and anti-Stokes fields $S,N$. The efficiency
of this process may be quite high, transferring up to 2-5 \% of the
incoming laser radiation to these new fields. Since the frequencies
of the generated fields are shifted by the ground-state hyperfine
splitting  ($6.8$~GHz for ${}^{87}$Rb) from the laser field, the
homodyne detector is not sensitiv to their fluctuations. However,
these Raman processes affect the atomic coherence and disturb the
correlation between the phase and the photon number of the laser
field.


Radiation trapping of spontaneous radiation~\cite{Novikova} is
another possible explanation why no vacuum squeezing was observed in
the experiment. Reabsorption of de-phased and de-polarized
spontaneous photons destroys atomic coherence. This process is
particularly important in optically dense atomic vapor, where
reabsorption probability is high: the negative effect of the
radiation trapping on ground-state coherence lifetime begins at
atomic density $\ge 10^{10} \mathrm{cm}^{-3}$. Additional decoherence
due to radiation trapping grows quickly with atomic density, and for
$N \simeq 5\times10^{11} \mathrm{cm}^{-3}$ becomes comparable to the
transient ground-state decoherence rate
$\gamma_{bc}$~\cite{Novikova}. As a result the coherent EIT medium
becomes more opaque, and this additional absorption reduces or
destroy vacuum squeezing. Please note that this effect is stronger
for the $D_2$ line due to the cycling transition $5S_{1/2}
F=2\rightarrow 5P_{3/2} F'=3$. Nonetheless, vacuum squeezing was
detected for the $D_2$ line in \cite{Lvov}, which is indirect
evidence that for that experiment the effect of atomic coherence was
minimal. We also note that the theory developed in \cite{Matsko} did
not account for radiation trapping.
%

Frequency modulation to amplitude modulation (FM-AM) conversion is
usually a significant noise source for experiments with diode lasers;
it is particularly important in atomic frequency standards and
magnetometers~\cite{Camparo}. Due to the relatively low quality
factor of the diode laser cavity, it has a wide phase noise spectrum
which is transferred into transmitted intensity fluctuations  after
traversing a resonant absorbing medium. In this experiment, however,
we used an extended cavity diode laser with a much lower phase noise
level. Thus we believe that FM-AM conversion is not the reason for
the observed quantum noise spectrum.
%

In conclusion, we studied the modification of the quantum noise of
linearly polarized laser radiation after interaction with a Rb vapor
cell. We observed a two order of magnitude enhancement of quantum
noise for certain phases of homodyne detector. The spectrum on this
excess noise has a step-like shape. The results presented here are
useful in development of atomic magetometers~\cite{Wynands} and
microwave frequency standards~\cite{Sergei} based on EIT.


Authors would like to thank A.\ B.\ Matsko for helpful discussions
and interest to this work, and E.\ A.\ Goldschmidt for her help with
the manuscript translation.
\def\baselinestretch{1.1}

\end{document}